\documentclass{article}

\def\xxinput#1{\input#1}

\xxinput{vsolj02.sty}

\usepackage{longtable}
\usepackage[dvipdfmx]{graphicx}
\usepackage[comma,colon]{natbib}

\usepackage[OT2,T1]{fontenc}

\def\cite{\citealt}
\setcitestyle{aysep={}}

\newcounter{author}
\def\altaffilmark#1{$^{#1}$}
\def\altaffiltext#1{$^{#1}$\,}

\def\authorcount#1#2{{\refstepcounter{author}\label{#1}
                     \altaffiltext{\ref{#1}}{#2}}}

\begin{document}

\begin{center}

\title{Z Cam star PY Per in SU UMa state?}

\author{
        Taichi~Kato\altaffilmark{\ref{affil:Kyoto}}
}
\email{tkato@kusastro.kyoto-u.ac.jp}

\authorcount{affil:Kyoto}{
     Department of Astronomy, Kyoto University, Sakyo-ku,
     Kyoto 606-8502, Japan}

\end{center}

\begin{abstract}
\xxinput{abst.inc}
\end{abstract}

   PY Per was discovered by \citet{hof66an289139} (S~9160)
as an RW Aur star (the type currently known as pre-main sequence
rapid irregular variables or the GCVS type ``IS'')
with a photographic range of 14--16.5~mag.
\citet{hof66an289139} made a remark that the star was
not particularly colored and that the amplitudes on Sonneberg
plates were 1.5--2~mag.  \citet{bon78bluevar2} systematically
studied high-latitude blue variables and found that
PY Per had diffuse (i.e. broad) hydrogen emission lines,
superposed on a blue continuum.  \citet{bon78bluevar2} classified
PY~Per as a possible dwarf nova.  \citet{zwi94CVspec1} obtained
a low-resolution spectrum and confirmed the finding
by \citet{bon78bluevar2}.  \citet{tay96arandamcaspyper}
obtained a radial-velocity study and found an orbital
period ($P_{\rm orb}$) of 0.15480(17)~d.  \citet{tay96arandamcaspyper}
pointed out that PY Per was very faint ($V \sim$19.8)
on an image in \citet{DownesCVatlas1}.

   Although this object has been known as a Z Cam star
and was relatively well observed by the American Association of
Variable Stars (AAVSO) members particularly between
2003 and 2017, the coverage has not been good as
in the past in recent years.  During the period of
intensive observations by the AAVSO, the object showed
both bright and faint states and the faint states
apparently lasted more than a year, particularly in
late 2008 to 2009.  During this faint state, there
was no detected outburst and the object can be classified
as a Z Cam+VY Scl star.  During bright states, the range
of variation was relatively small (14.0--17.0), which probably
corresponded to the state observed by \citet{hof66an289139}.
The faint record on an image in \citet{tay96arandamcaspyper}
was not that faint and was brighter than 19.0 compared
to Gaia EDR3 magnitudes of nearby stars \citep{GaiaEDR3}.
The state when the image in \citet{DownesCVatlas1} was
obtained was probably similar to faint states
recently observed.

\begin{figure*}
\begin{center}
\includegraphics[width=16cm]{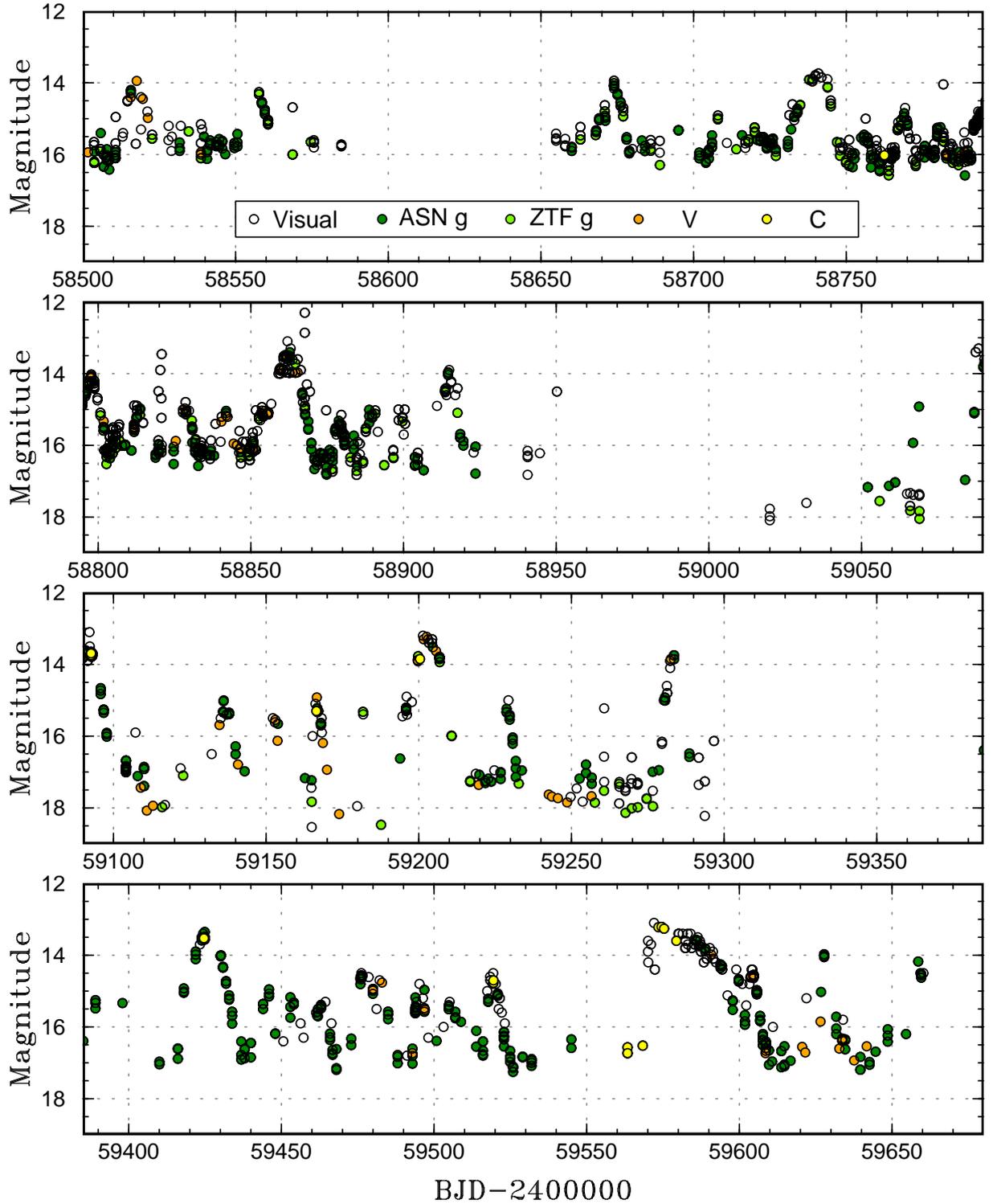}
\caption{
  Long-term light curve of PY Per using
  VSOLJ, VSNET, AAVSO, ASAS-SN and ZTF observations.
  The object showed low-amplitude outbursts before 2020
  (the first and second panels).  The outburst pattern
  changed since then and the behavior in the bottom panel
  resembles that of an SU UMa star in that long, bright
  outbursts occur in addition to short, faint outbursts.
}
\label{fig:pyperlc}
\end{center}
\end{figure*}

\begin{figure*}
\begin{center}
\includegraphics[width=16cm]{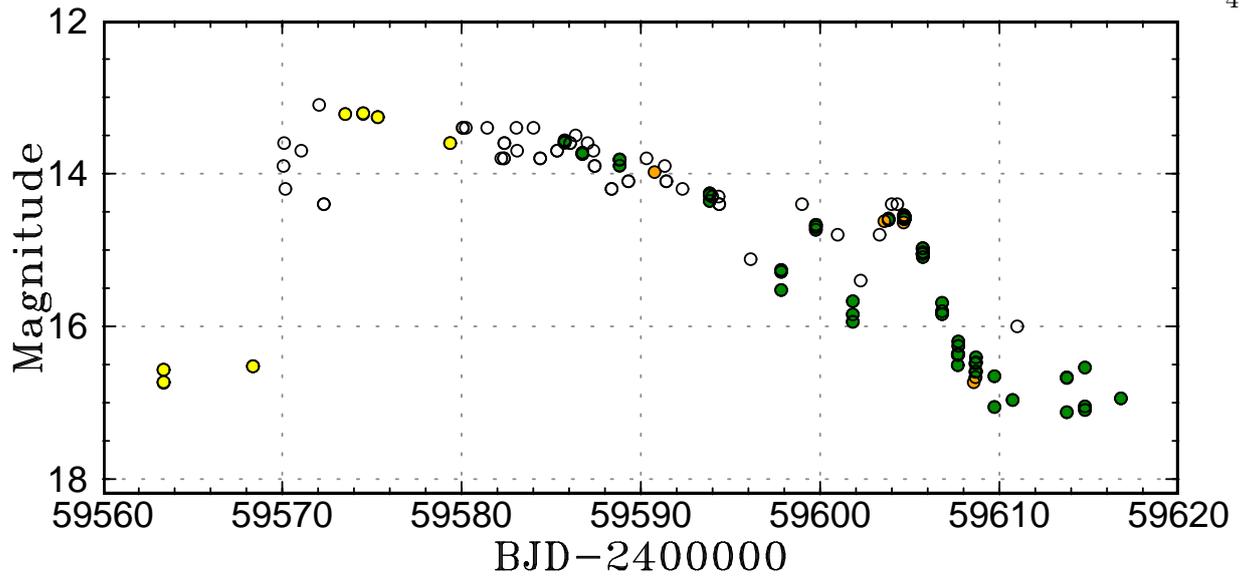}
\caption{Enlargement of the 2021 December--2022 January
  long and bright outburst of PY Per.  There were apparently two
  rebrightenings during the fading phase of the outburst.
  The symbols are the same as in figure \ref{fig:pyperlc}.
}
\label{fig:pyperlc2}
\end{center}
\end{figure*}

   I noticed that the outburst behavior changed dramatically
since 2020, when relatively regularly recurring
long, bright outbursts with cycles of 110--160~d
and short, faint outbursts between them dominated
in the light curve (figure \ref{fig:pyperlc}).
I used the All-Sky Automated Survey for Supernovae (ASAS-SN)
Sky Patrol data \citep{ASASSN,koc17ASASSNLC},
the Zwicky Transient Facility (ZTF: \cite{ZTF})
data\footnote{
   The ZTF data can be obtained from IRSA
$<$https://irsa.ipac.caltech.edu/Missions/ztf.html$>$
using the interface
$<$https://irsa.ipac.caltech.edu/docs/program\_interface/ztf\_api.html$>$
or using a wrapper of the above IRSA API
$<$https://github.com/MickaelRigault/ztfquery$>$.
}, observations by AAVSO International Database\footnote{
   $<$http://www.aavso.org/data-download$>$.
}, Variable Star Observers League
in Japan (VSOLJ) and VSNET \citep{VSNET}.
The pattern even looks like that of an SU UMa star.
The long outburst in 2021 December--2022 January
gradually faded and had a duration more than 25~d, which
was unexpectedly long, and even had two rebrightenings
on the fading phase (figure \ref{fig:pyperlc2}).
This outburst particularly looks like a superoutburst.

\begin{figure*}
  \begin{center}
    \includegraphics[width=16cm]{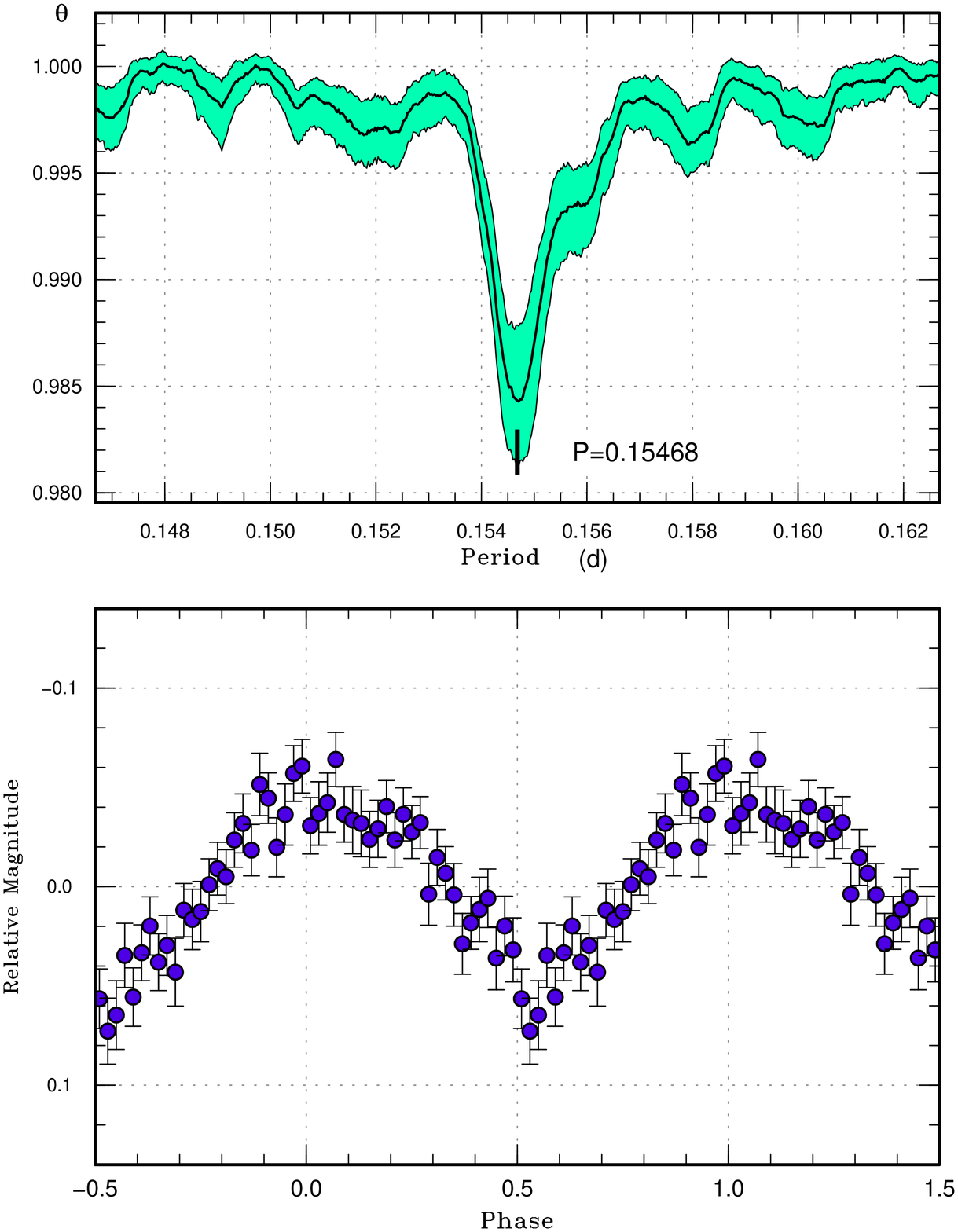}
  \end{center}
  \caption{Period analysis of the TESS data.
  (Upper): We analyzed 100 samples which randomly contain 50\% of
  observations, and performed the PDM analysis for these samples.
  The bootstrap result is shown as a form of 90\% confidence intervals
  in the resultant PDM $\theta$ statistics.
  (Lower): Orbital variation.
  }
  \label{fig:pdm}
\end{figure*}

   I also analyzed Transiting Exoplanet Survey Satellite (TESS)
observations.\footnote{
  $<$https://tess.mit.edu/observations/$>$.
}  The full light-curve
is available at the Mikulski Archive for Space Telescope
(MAST\footnote{
  $<$http://archive.stsci.edu/$>$.
}).  The TESS observations were obtained in 2019 November
(BJD 2458790--2458814), when the amplitudes of dwarf nova-type
variations were small.
The $P_{\rm orb}$ was determined to be 0.15468(5)~d by using
the Phase Dispersion Minimization (PDM, \cite{PDM}) method
after removing long-term trends by locally-weighted polynomial regression
(LOWESS: \cite{LOWESS}) (figure \ref{fig:pdm}).
The errors of periods by the PDM method were
estimated by the methods of \citet{fer89error} and \citet{Pdot2}.
This result confirmed and slightly
refined the value in \citet{tay96arandamcaspyper}.
There was no hint of positive or negative superhumps.

  If PY Per indeed showed a superoutburst, it is unusual for
an object with $P_{\rm orb}$ of 0.15468(5)~d.
There is, however, a confirmed case of BO Cet with
$P_{\rm orb}$=0.1398~d \citep{kat21bocet}.  A suspected
case of ASASSN-14ho [$P_{\rm orb}$=0.24315(10)~d,
\cite{gas19asassn14hov1062cyg}] with multiple rebrightenings
has also been reported \citep{kat20asassn14ho}.
Although there was apparently no time-resolved photometry
during the 2021 December--2022 January outburst and
the object has already gone into solar conjunction,
I would like to call attention to this object to see
whether this state continues when the object emerges
in the morning sky and if it is the case, time-resolved
photometry during a long, bright outburst is desired.
It is known that SU UMa stars with long $P_{\rm orb}$
can have three types of outbursts
(normal, long normal and superoutbursts) as in TU Men
\citep{war95suuma,bat00tumen}, NY Ser \citep{pav14nyser}
and V1006 Cyg \citep{kat16v1006cyg,pav18v1006cyg}.
The difference in the durations of long outbursts
in PY Per after 2020 may reflect different types
of outbursts and only the 2021 December--2022 January
might have been a genuine superoutburst.

\section*{Acknowledgements}

This work was supported by JSPS KAKENHI Grant Number 21K03616.
The author is grateful to the ASAS-SN and ZTF teams
for making their data available to the public.
I am grateful to VSOLJ, AAVSO and VSNET observers for
reporting observations and to Naoto Kojiguchi for
helping downloading the ZTF data.

Based on observations obtained with the Samuel Oschin 48-inch
Telescope at the Palomar Observatory as part of
the Zwicky Transient Facility project. ZTF is supported by
the National Science Foundation under Grant No. AST-1440341
and a collaboration including Caltech, IPAC, 
the Weizmann Institute for Science, the Oskar Klein Center
at Stockholm University, the University of Maryland,
the University of Washington, Deutsches Elektronen-Synchrotron
and Humboldt University, Los Alamos National Laboratories, 
the TANGO Consortium of Taiwan, the University of 
Wisconsin at Milwaukee, and Lawrence Berkeley National Laboratories.
Operations are conducted by COO, IPAC, and UW.

The ztfquery code was funded by the European Research Council
(ERC) under the European Union's Horizon 2020 research and 
innovation programme (grant agreement n$^{\circ}$759194
-- USNAC, PI: Rigault).

\section*{List of objects in this paper}
\xxinput{objlist.inc}

\section*{References}

We provide two forms of the references section (for ADS
and as published) so that the references can be easily
incorporated into ADS.

\renewcommand\refname{\textbf{References (for ADS)}}

\newcommand{\noop}[1]{}\newcommand{\hyphalt}{-}

\xxinput{pyperaph.bbl}

\renewcommand\refname{\textbf{References (as published)}}
\xxinput{pyper.bbl.vsolj}

\end{document}